\documentclass[12pt]{article}

\usepackage{amssymb}

\def\be{\begin{equation}}
\def\ee{\end{equation}}
\def\bea{\begin{eqnarray}}
\def\eea{\end{eqnarray}}
\def\d{\partial}
\def\D{\nabla}

\def\g{\gamma}

\def\icv{\nu_1,...,\nu_d }
\def\A{A(X)_{\icv}}
\def\FA{F(X)_{\m\icv}}
\def\pxd{\prod_{a=1}^{d}\d_{a}X^{\nu_a}}
\def\pxa{\partial_{a}X^{\mu}}

\def\ppi{\partial_{\mu}\phi^i (X)}
\def\ppj{\partial_{\nu}\phi^j (X)}
\def\pdt{\prod_{i=1}^{\td} \delta (\phi^{i}(X))}

\def\ppd{\prod_{j=1}^{\td} \d_{\m_{j}}\phi^{j}(X)}
\def\ppdi{\prod_{j=1, j\neq i}^{\td} \d_{\m_{j}}\phi^{j}(X)}

\def\ic{^{\rho}_{\mu\nu}}

\def\G{G_{00}}  
\def\S{\Sigma}

\def\td{\tilde{d}}

\def\ve{\varepsilon}

\def\e{\epsilon}

\def\g{\gamma}
\def\G{\Gamma}

\def\m{\mu}
\def\n{\nu}

\def\D{\nabla}

\def\O{\Omega}

\def\S{\Sigma}

\def\sqg{\sqrt{\g}}
\def\sqp{\sqrt{G\, h}}
\def\sqqp{\sqrt{Q^{+}\, h^{+}}}

\begin{document}
\begin{titlepage}

\begin{flushright}
IFT-UAM/CSIC-98-6\\
hep-th/98mmnnn \\
\end{flushright}

\vspace{1.5cm}
\begin{center}
\large{{\bf A DUAL FORMULATION 
FOR (p and D)-BRANES VIA TARGET-SPACE 
FIELD THEORY } } \\
\vspace{1.0cm}

\large{\bf {Javier Borlaf }} \\
\vspace{2mm}

Instituto de F\'isica Te\'orica, CXVI\\
Unidad de Investigaci\'on Asociada \\al 
Centro de F\'{i}sica Miguel Catal\'an
(C.S.I.C.)\\
and\\
Departamento de F\'isica Te\'orica,CXI\\
Universidad Aut\'onoma, 28049 Madrid, Spain \\
javier@delta.ft.uam.es,

\end{center}
\vspace{1.5cm}

\begin{abstract}
It is shown how some field theories in the
target-space induce the splitting of 
the space-time into a continuous of branes,
which can be p-branes or D-branes depending 
on what the field theory it is. 
The basic symmetry underlying the construction
is used to build an invariant action, which
is proved to be off-shell identical to the
p-brane (D-brane) action. The coupling with
the abelian ($p+1$)-form in this formulation
it is also found. While the classical brane's
embedding couple to the field strenght, 
the classical fields couple with its dual 
(in the Hodge sense), therefore providing an 
explicit electric-magnetic duality.
Finally, the generic role of the underlying symmetry
in the connection between the target-space theory 
and the world-volume one, is completely elucidated.
\end{abstract}

\vspace{15mm}

\noindent

\end{titlepage}

\section{Introduction}
\setcounter{equation}{0}
Strings are the natural extension of point particles and in the last
twenty five years they have proved to be a privileged framework
for broaching the unification of all known interactions,
including specially the elusive gravity \cite{schwarz1}.
Despite of the apparent 
crucial feature of the bidimensionality, mainly conformal invariance,
some people considered that other extended objects, called branes,
could play a role in this history (for two excellent review see 
\cite{duff1}). 
This attempt deflates by the belief that kappa symmetry, 
of decisive importance on building superparticles and 
superstrings a la Green-Schwarz \cite{schwarz2}, 
is not generalizable to higher
dimensional branes. Even more, the supermembrane seems to be
quantically unstable. However, some time later, kappa-symmetric 
actions appeared for some branes in different dimensions. 
Aside from thoses advances, string theory 
centers the main part of the researcher effort. 
But, fortunatelly,  this is not the end of the brane's story.

\hfill

In the same way that particles couple to the electromagnetic field, 
which is a one form in the target-space, branes of dimension d, 
called (d-1)-branes, couple to d-forms.  The effective action 
of the heterotic string presents a dual formulation ( in the forms sense)
in terms of a six form. This fact suggested  the existence of an
S-dual formulation of the heterotic string in terms of a 
fundamental five-brane. The strong coupling of the heterotic
string would be better described in the weak regime of the five-brane. 
This was called the string/five-brane duality 
conjectured \cite{duff3}. From now, the role of branes in string theory 
is just starting. 
The relation between the p-forms of the effective actions  and 
the (p-1)-branes has been supported by the obtention of the corresponding
elementary and solitonic solutions in every supergravity in ten dimensions.
In particular, in \cite{strominger} 
was shown how the heterotic effective field theory 
admits the five-brane as a soliton solution,then  backing
the conjectured string/five-brane duality. 

\hfill

But extended objects burst into string theory in another very exotic and 
surprising way by means of the Dirichlet-branes \cite{polchinski}.
These are the hypersurfaces
in which the strings with Dirichlet boundary conditions rest their
endpoints. 
Consistency requires that these branes must be promoted to dynamical
objects
carrying the Ramond-Ramond charges needed by duality.
Therefore they are automatically included in the spectrum. 

\hfill

The discovery of the connection between the different compactifications 
of sugras in ten dimensions and the supergravity in eleven dimensions
supported the idea of an underlying fundamental 11-dimensional theory,
called, the M-theory \cite{witten}.
Its low energy regime, the eleven dimensional
supergravity, 
contains a three-form. Maybe there is a unique fundamental brane, the
membrane in eleven dimensions. 

\hfill

As we can see, although string theory seems to be the unique starting point
of all resulting scenarios, it is not a crazy idea that the underlying
model could have as elemental consituent (s) other brane(s) than strings.

\hfill

The usual description of  a d-dimensional extended object $\S$ 
starts with its embeddeding
in the D-dimensional space-time, i.e., $X^{\m}(\ve^a)$, where $\{\ve^a\}$ 
with $a=1\,to\,d$ parametrize the (d-1)-brane. The dynamic is a direct
extension to the corresponding with the point particle one : the
hypervolume
swept up by the classical configuration in its time evolution is an
stationary 
point of the action. Then, the coupling with the metric is the Nambu-Goto 
type action :

\be
\label{p1}
S_p^{G}= \int_{\S} d^d \ve  \sqrt{\g}
\ee

\hfill

where $\g_{ab}=G_{\m\n}(X)\d_{a}X^{\m}\d_{b}X^{\n}$ is the metric
induced in the brane and  $\gamma$ its determinant \footnote{We use
euclidean
metrics through the paper.}. The coupling with the abelian electromagnetic
field,
i.e. , the d-form $\A$, follows the pattern as in the point particle case :

\be
\label{p1bis}
S_p^{A}= \int_{\S} d^d \ve   \A \pxd 
\ee

\hfill

Obviously, both actions are invariant under reparametrizations in the
world-volume $\ve^a  \rightarrow \ve^{'a}(\ve^{b})$ and so it is the
dynamic, obtained from the whole action $S_{p}= S_p^{G}+ S_p^{A}$ 
\footnote{${\delta S \over \delta X^{\m}}\equiv
\d_{a}({\d {\cal L} \over \d \d_{a}X^{\m}})-
{\d {\cal L}\over \d X^{\m}}$.} :

\be
\label{p2}
{\delta S_p \over \delta X^{\m}} =
\sqg  P^{\S}_{\m\n}\Delta X^{\n}  -  \FA \pxd =0
\ee

\hfill

where the laplacian $\Delta X^{\rho}$ is conveniently defined as :

\be
\label{p3}
\Delta X^{\rho}= \g^{ab}(\d_{a}\d_{b}X^{\rho} + \G\ic
\d_{a}X^{\m}\d_{b}X^{\n})
\ee

\hfill

which is manifestly covariant under diffeomorphisms in the 
space-time  $X^{\m} \rightarrow X^{'\m}(X^{\n})$, but no 
under reparametrizations in the world-volume. It is just the 
presence of the ortogonal (to the brane) proyector 
$ P^{\S \,\,\m\n}\equiv G^{\m\n} - U^{\S\,\,\m\n}$, with 
$ U^{\S\,\,\m\n}= \g^{ab}\d_{a}X^{\m}\d_{b}X^{\m}$, 
which guarantees the invariance of  (\ref{p2}) under 
diffeomorphisms in the world-volume. The field strength 
for the p-form is $\FA\equiv \d_{\m}\A -\d_{\n_{1}}A(X)_{\m\n_{2},...,
\n_{d}}- ...- \d_{\n_{d}}A(X)_{\n_{1}\n_{2},...,
\m}$.

\hfill

The D-branes are extended objects of a different class than 
p-branes. They appear in the context of open strings with 
Dirichlet boundary conditions, but, through this paper,
we are only interested in two features : firstly, the coupling
with the metric is given by the substitution $G_{\m\n}\rightarrow
G_{\m\n}+B_{\m\n}$ on (\ref{p1}), where $B_{\m\n}=-B_{\n\m}$ is the 
Neveu-Schwarz (NS-NS) two form recalling the stringy origin 
for this brane. 
This is called a DBI-(Dirichlet-Born-Infeld)-Action \cite{leigh}.
Second, the electromagnetic term is the same type as in (\ref{p1bis}),
but the d-form is now a composite one, given by a combination of the
Ramond-Ramond (R-R) forms and the NS-NS two form \cite{dbrane}. 
The presence of the two-form in the DBI-action 
makes the equations very involved, so we let them to the 
appendix.

\hfill

The purpose of this work consists on showing the existence of a 
reformulation for the dynamics of p-branes, D-branes and 
any other branes  
in terms of scalar fields in the target-space. The idea comes from 
\cite{sugamoto, yabuki} where this was made in a few very simple examples.
Here we will give the basic symmetry which shall serve to guide us for
building
the field equations for our scalar fields, no matter the dimension D of the
space-time, its geometry, and the dimension d of the brane is. 
This symmetry, which basically
consists on arbitrary redefinitions of the scalar fields, will allow us to
interpret those fields as defining a hypersurface which will result to be
the p-brane ,or D-brane, depending on the field theory we are dealing with.
In this formulation, the coupling
with the electromagnetic d-form is also given. This will serve to see that 
both formulations are classically related by electric-magnetic duality. 
Even more, certain ambiguities on defining an action
for the field theory having the right symmetry, are solved in a
natural way by demanding its correspondence with branes. 
After that, the off-shell identity between this action and the p-brane
(or D-brane) action is shown, totally supporting the construction. 
Finally, the power of the symmetry under field redefinitions is used
to justify the intrinsic relation between the target-space degrees 
of freedom and  the world-volume ones, no matter the specific form of the
action is. The rule giving the target-space field theory once 
the brane one is known, is also found. In this general context, the 
classical equivalence it is shown in detail.   
\hfill

In section 2, the basic symmetry is introduced and the simplest 
covariant field equations are built. Section 3 shows that the branes
induced by those equations are classical p-branes in absence 
of electromagnetic field. In section 4, the problems on building 
the right action are solved demanding its brane origin and 
the invariance under field redefinitions. Then, the off-shell duality
with the p-brane action is shown. In section 5, the electromagnetic
contribution is found in this formulation, giving the p-brane's one
after dualization. The classical electric-magnetic duality is manifest.
A simple example is worked out. In section 6, the invariance under
field redefinitions is picked  out as the feature responsible for 
the exchange of degrees of freedom between both formulations.
The rule assigning the target space lagrangian in terms of the
world-volume brane one is explicitely given. The explicit 
classical equivalence it is shown.
Section 7 summarize the D-brane case. We end with the conclusions. 
\section {The Classical Model}
The idea consists \cite{sugamoto, yabuki}
on describing an extended object $\S$ by means of 
usual fields, i.e., fields defined in the target-space ${\cal M}$,but
not in an auxiliar manifold (world-volume). We start with a set of 
$\td$ scalar fields $\phi^i (X)\,\,$i$=1\,$to$\,\td=D-d$. Our hypersurface 
$\S$ is defined as the collection of points in the space-time where 
the whole set of  fields vanish :

\be
\label{d1}
\S \equiv \{ X \in M  \setminus  \phi^i (X) = 0 \,\,\forall \,\,i=1 \,to\,
\td \}
\ee

\hfill

This is a very simple an usual way to represent hypersurfaces.
Now, the main problem is to build a dynamic for those fields 
which allow us to interpret them as representing the extended 
object (\ref{d1}). The key observation of this work is that 
arbitrary redefinition of fields

\be
\label{d2}
\phi^i (X) \longrightarrow \phi^{'\,i}(\phi^j (X))
\ee

\hfill

must be a symmetry of the field equations always than
det${\d \phi^{'i} \over \d \phi^{j}} \neq 0$. In that way
the hypersurfaces  $\Sigma^{'}$ and $\Sigma$,
defined as $\phi^{'}(X)=0$ and $\phi(X)=0$, are the same. 
In this formalism, the invariance under (\ref{d2}) will become 
equivalent to the invariance under
reparametrizations in the worl-volume in the $X^{\m}(\ve^a)$
description. It is just the invariance under (\ref{d2}) which
guide us to build the right equations. Let us see how this works.

The basic first-derivative 
target-space scalar we get is 

\be
\label{d3}
h^{ij}(X) = G(X)^{\m\n}\ppi \ppj
\ee 

\hfill

Because the $\d \phi^{i}$ are ortogonal to the hypersurfaces 
$\phi = constant$, $h_{ij}$ can be viewed as a sort of transverse
metric. Moreover, is a $(2,0)$ tensor under field redefinitions :

\be
\label{d4}
h^{'\,ij}= {\d \phi^{' i}\over \d \phi^{k}}{\d \phi^{' j}\over \d
\phi^{l}} h^{kl}
\ee

\hfill

Now, for building the dynamics compatible with (\ref{d2}),
we just need to realize the existence of a natural connection
for our symmetry transformation (\ref{d2}):

\be
\label{d5} 
(\O_{\m})_{i}^{j} \equiv \O_{\m\,\,i}^{j} =
h_{ik}\d^{\n}\phi^{k}\D_{\m}\d_{\n}\phi^{j}
\ee

\hfill 

where $\D$ is the target-space covariant derivation built with
the corresponding connection $\G_{\m\n}^{\rho}$ \footnote
{Through this work, $\Gamma_{\m\n}^{\rho}$ stands for the
Levi-Civita connection and $\D_{\m}$ its corresponding covariant 
derivation.} and  $h_{ij}$ satisfies
$h_{ik}h^{kj}=\delta^{i}_{j}$.

In matrix notation, the $\O_{\m}$ transformation under field 
redefinitions is

\be
\label{d6}
\O^{'}_{\m}=  {\d \phi^{'}\over \d \phi}\O_{\m} {\d \phi \over \d
\phi^{'}}
+ \d_{\m}({\d \phi^{'}\over \d \phi}) {\d \phi \over \d \phi^{'}}
\ee

\hfill

that is, $\O$ is a real connection. Therefore, we can give covariant
derivatives
under (\ref{d2}) using the basic ones for the 
$(1,0)$ ($V^i$) and $(0,1)$ ($V_{j}$) tensors :

\bea
\label{d62}
&V^{'i}=  {\d \phi^{' i}\over \d \phi^{k}} V^{k} \rightarrow
D_{\m}V^{i}\equiv \d_{\m}V^{i}-\O_{\m\,j}^{i}V^{j}\rightarrow
D_{\m}^{'}V^{'i}= {\d \phi^{' i}\over \d \phi^{k}} D_{\m}V^{k}\nonumber\\
&V^{'}_{i}=  {\d \phi^{ j}\over \d \phi^{'i}} V_{j}
\rightarrow D_{\m}V_{i}\equiv \d_{\m}V_{i}+ \O_{\m\,i}^{j}V_{j}
\rightarrow D_{\m}^{'}V_{i}^{'}= {\d \phi^{ j}\over \d \phi^{'i}}
D_{\m}V_{j}
\eea

\hfill

In particular, the "transverse metric" is covariantly constant :
$D_{\m} h^{ij} = \d_{\m} h^{ij} - \O_{\m\,k}^{i} h^{kj}
-\O_{\m\,k}^{j} h^{ik}= 0$.

\hfill

Once the connection has been established, the simplest dynamic 
we can build is :

\bea
\label{d7}
&D_{\m}\d^{\m}\phi^{i}= G^{\m\n}D_{\m}\d_{\n}\phi^{i} \nonumber\\
&= G^{\m\n}(\d_{\m}\d_{\n}\phi^{i} -
\O_{\m\, j}^{i}\d_{\n}\phi^{j}- \G_{\m\n}^{\sigma}\d_{\sigma}\phi^{i})=0
\eea

\hfill

where the target-space connection is also present to guarantee the 
invariance under diffeomorphisms on {\cal M}. It will be very useful
for studying their relation with the p-branes to rewrite (\ref{d7}) as :

\be
\label{d8}
D_{\m}\d^{\m}\phi^{i}=U^{\m\n}\D_{\m}\d_{\n}\phi^i=
U^{\m\n}(\d_{\m}\d_{\n}\phi^{i} -
 \G_{\m\n}^{\sigma}\d_{\sigma}\phi^{i})=0
\ee

\hfill

in terms of the proyector  $U_{\m\n}= G_{\m\n}-P_{\m\n}$
with $P_{\m\n}\equiv h_{ij}\ppi \ppj$, satisfying the properties
$U^{\,\,\,\m}_{\n}\ppi =0$ and $ U^{\,\,\,\n}_{\m} U^{\,\,\,\rho}_{\n}=
U^{\,\,\,\rho}_{\m}$ \footnote{The indices are lowered (raised ) with the
metric 
$G_{\m\n}$ ($G^{\m\n}$)}.

It is just the contraction with the projector $U$ instead of the
metric $G$ in (\ref{d8}) which makes that laplacian invariant
under field redefinitions, at the same time that the own dependence 
of $U$ on the fields makes the equations highly non-linear.

\hfill

\section{The Duality with p-branes}

Now we want to see what are the conditions imposed by the field equations 
(\ref{d8}) in one embedding $X^{\m}(\ve^{a})$ for the hypersurface $\S$
defined in (\ref{d1}). Therefore, the fundamental relation between the
scalar
fields and the embedding is :

\be
\label{d9}
\phi^{i}(X^{\m}(\ve^a))=0
\ee

\hfill

which must be understood as a  set of equations for the embedding 
once the scalar fields are given  \footnote{A circle of radius 1 
in two dimensions is described by means of the relation
$\phi (X, Y) =X^2 + Y^2 -1 =0$. 
A solution for (\ref{d9}) is the embedding 
$X=\cos \theta $ and $Y=\sin \theta $, 
which parametrizes the surface (curve) $\phi(X,Y)=0$
with $\theta $ running from  $0$ to $2 \pi$,
and both points identified.}.

Taking derivatives with respect to $\ve^a$ :

\be
\label{d10}
{\d X^{\m}\over \d \ve^a}\d_{\m}\phi^{i}(X(\ve))=0
\ee

\hfill

which immediately implies that $U^{\S}$ annhilates $\ppi$
when the last one is evaluated on $\S$, and, at the same time 
$U$ evaluated over $\S$ is the identity for  $\pxa$
\footnote{From now, when an object, say $\eta (X^\m)$, is 
evaluated on the hypersurface $\S$, we write $(\eta)_{\S}$.} :

\bea
\label{d11}
&U^{\S\,\,\m}_{\n}(\ppi)_{\S}=0\\
&(U^{\m}_{\n})_{\S}\d_{a}X^{\n}= \pxa
\eea

\hfill

This allow us to identify $U$ evaluated on $\S$ with $U^{\S}$

\be
\label{d12}
U^{\S\,\,\m\n}=(U^{\m\n})_{\S}
\ee

\hfill

Therefore, we can relate the field equations (\ref{d8}) with the 
p-brane equations (\ref{p2}) for the embedding, taking account
(\ref{d10}) and (\ref{d12}), in the following way 
\footnote{Here without the d-form.}:

\bea
\label{d13}
&(D_{\m}\d^{\m}\phi^{i})_{\S}= 
(U^{\m\n}\D_{\m}\d_{\n}\phi^i )_{\S}=\nonumber\\
&\g^{ab}\d_{a}X^{\m}\d_{b}X^{\n}(\d_{\m}\d_{\n}\phi^{i} -
 \G_{\m\n}^{\sigma}\d_{\sigma}\phi^{i})=\nonumber\\
&\g^{ab}{\d \,\,\over \d \ve^{a}}\{{\d X^{\m}\over \d
\ve^b}\d_{\m}\phi^{i}(X(\ve))\}
- \d_{\m}\phi^{i}\g^{ab}\d_{a}\d_{b}X^{\m}- \g^{ab} \G\ic
\d_{\rho}\phi^{i} \d_{a}X^{\m}\d_{b}X^{\n}=\nonumber\\
&=-(\ppi)_{\S}\Delta X^{\m}
\eea

\hfill

But using  $P_{\m\n}\equiv h_{ij}\ppi
\ppj$, we finally obtain :

\be
\label{d14}
(h_{ij}\ppj\,D_{\m}\d^{\m}\phi^{i})_{\S}= 
-P^{\S}_{\n\m}\Delta X^{\m}=0
\ee

\hfill

i.e., the equations for the scalar fields $\phi^i (X)$ evaluated
over the hypersurface $\S$ defined as $\phi^i (X)=0$ for all $i$,
imply the p-brane equations (\ref{p2})
for any embedding $X^{\m}(\ve^a)$ of 
$\S$. 
Following the same criteria it is easy to show that the converse is true :
the p-brane equations for the embedding of some hypersurface $\S$ imply
the field equations (\ref{d8}) for any set of scalar fields satisfying
(\ref{d9}).

\hfill

\be
\label{d14bis}
(D_{\m}\d^{\m}\phi^{i})_{\S}= 0 \Longleftrightarrow
P^{\S}_{\n\m}\Delta X^{\m}=0
\ee

\hfill

In addition to the symmetries already mentioned, the dynamic for 
the scalar fields presents a trivial one under global translations :

\be
\label{d14bisbis}
\phi^i (X^{\m}) \longrightarrow \phi^i (X^{\m}) + C^i
\ee

\hfill

Therefore every solution is not representing a unique p-brane, but
the splitting of the entire manifold ${\cal M}$ into a continuous 
of p-branes $ \phi^i (X^{\m})= C^i $ every one of them parametrized
by the set $C^i$ from $i=1 \,\,to\,\, \td$. It is a fluid of classical
p-branes.

\hfill

As a very simple example of this reformulation, it can 
be easily shown that, in flat space-time, the scalar fields 

\be
\label{d14bisbisbis}
\phi^i (X)= \phi^i ( A^{j}_{\m}X^{\m})
\ee

\hfill

are always a solution of the 
field equations (\ref{d8})
for any functional form 
$\phi^{i}(Y^j)$, such that
$(det {\d \phi^i \over 
\d Y^j }\neq 0)$
provided that the constant matrix  
$A^{i}_{\m}$
satisfies 
$det( A^{i}_{\m} A^{j}_{\m}) 
\neq 0$.
The last condition is equivalent to 
the requirement that
$ A^{i}_{\m}X^{\m}= C^i$ 
define a $d$-hyperplane,
which is the obvious d-brane in flat space.


\section{The Elementary Action and the off-shell Duality}

Up to now we have been able to build a dynamic for the scalar
fields based upon the symmetry guaranteeing their interpretation
as representing an extended object (or a continuous of them).
Moreover we have proved that those extended objects naturally
induced by the fields, are p-branes. Now we are in searching of
the minimal action principle reproducing the dynamic we have
found. Even more, we are interested in chequing the off-shell
equivalence, i.e., the relation between that action and the p-brane
action (\ref{p1}), using (\ref{d9}) as a kinematic relation but 
without the use of the equations of motion. 

\hfill

At first sight, we are in trouble because we find no first-derivative 
invariants under both, target-space diffeomorphismes and field
redefinitions.
However, the lagrangian density  ${\cal L}= \sqp$, with $G=det(G_{\m\n})$
and $h \equiv det(h^{ij})$ has the right variational derivative
\footnote{${\delta S\over \delta \phi^i}\equiv
\d_{\m}({\d {\cal L}\over \d \d_{\m}\phi^i})$.}:

\be
\label{d15}
{\delta \sqrt{G\,h} \over \delta \phi^{i}} =
\sqrt{G\,h} h_{ij}U^{\m\n}\D_{\m}\d_{\n}\phi^{j}=
\sqrt{G\,h} h_{ij}D_{\m}\d^{\m}\phi^{j}
\ee

\hfill

Although the corresponding action is invariant under target-space
diffeomorphisms, it is not the case for the redefinition of fields :

\be
\label{d16}
(\sqrt{G\,h})^{'}=\mid det( {\d \phi^{'}\over \d \phi})\mid 
\sqrt{G\,h}
\ee

In \cite{fairlie} it is shown how this kind of transformation is intimately
related with the non-uniqueness of the lagrangian giving rise to 
the field equations. In our case, the identity

\be
\label{d172}
\ppi {\d \sqp \over \d (\d_{\m}\phi^{j})}= 
\delta_{j}^{i}\sqp
\ee

\hfill

ensures that the lagrangian ${\cal L}_{F}= F(\phi){\cal L}=F(\phi)\sqp$
produces the same field equations independent of the arbitrary function
$F(\phi^1 ,...,\phi^{\td})$, which therefore acts as "constant" in the
variational derivative :

\be
\label{d17}
{\delta (F(\phi) \sqrt{G\,h}) \over \delta \phi^{i}}=
F(\phi) {\delta \sqrt{G\,\,h} \over \delta \phi^{i}}
\ee

\hfill

If we want to describe an elementary object, instead of a fluid
of them, the solution to the lagrangian ambiguity comes from
restricting the integration just to the elementary object $\S$.
The action is fixed to be :

\be
\label{d18}
S_{\phi}^{G}=\int_{\cal M}d^{D}X \,\sqp \,\pdt 
\ee

\hfill

Therefore, the arbitrarieness in the function 
$F(\phi^1 ,...,\phi^{\td})$ is trivial
 ($(F(\phi^1 ,...,\phi^{\td}))\vert_{\S}= F(0,0,...,0)$) and 
moreover, the presence of the delta functions makes the action 
invariant under field redefinitions, due to the identity 

\be
\label{d19}
\mid det( {\d \phi^{'}\over \d \phi}) \mid \prod_{i=1}^{\td} \delta
(\phi^{'i}(\phi^{j}(X)))
=\pdt
\ee

\hfill

that, of course, 
it must be understood within the integral. 
Then, the Dirac's deltas transformation 
compensates the one
for $\sqp$, and the action 
$S_{\phi}^{G}$
is now invariant under field redefinitions.

\hfill

The next step consist on showing the off-shell identity between the 
action (\ref{d18}) and the p-brane action (\ref{p1}). For that, we 
solve the set of equations  $\phi^i (X^{\m})=0$ 
defining $\S$ splitting locally the coordinates 
$X^{\m}= \{X^{a}, X^{i}\}$ ($i=1$ to
$\td$ and $a=\td + 1$ to $D$)  to write the solution in the way
$X^{i}=X^{i}(X^{a})$
\footnote{$ det( {\d \phi^{i}\over \d X^{j}})\vert_{\S}
\neq 0 $ in this splitting.}. 
Of course, this splitting guarantees
that we can choose locally the gauge  $X^a = \ve^a $ in the world-volume.
In this gauge (\ref{d10}) implies
$(\d_{a}\phi^i )_{\S}=-\d_{a}X^{j}(\d_{j}\phi^{i})_{\S}$
which allow to get the fundamental relation :

\be
\label{d19bis}
(det( {\d X^{i}\over \d \phi^{j}})^2\,G\,h)_{\S} = \gamma
\ee

\hfill

obtained with the help of the identity

\bea
\label{d19bis2}
&det\{1+A_{2}B_{1}+B_{2}A_{1}+A_{2}(1+B_{1}B_{2})A_{1}\}=
\nonumber\\
&det\{1+B_{1}A_{2}+A_{1}B_{2}+A_{1}(1+B_{2}B_{1})A_{2}\}=
\eea

\hfill

where $A_{1}, B_{1}$ are arbitrary $d \times \td$ matrices 
and  $A_{2}, B_{2}$ are arbitrary $\td \times d$ matrices
\footnote{This identity can be obtained from
the determinant of the auxiliar $D\times D$ matrix

\[  \left( \begin{array}{cc}
 (1+ B_{1}B_{2})^{-1}& - ( A_{1}+ (1+ B_{1}B_{2})^{-1} B_{1})\\
( A_{2}+ B_{2} (1+ B_{1}B_{2})^{-1}) & (1+B_{2}B_{1})^{-1}
\end{array}
\right) \]
}.

\hfill

Now, using (\ref{d19}), the integration in the $X^{i}$ can be made
resulting in the $X^{i}=X^{i}(X^{a})$  restriction and a jacobian 
$\mid det( {\d X^{i}\over \d \phi^{j}})\mid $:

\bea
\label{d20}
&S_{\phi}^{G}=\int  \,d^d X (\int \pdt dX^{i} \sqp )=\nonumber\\
&\int d^d X (\mid det({\d X^{i}\over
\d \phi^{j}})\mid \sqp)\vert_{X^{i}=X^{i}(X^{a})}
\nonumber\\
&= \int d^d X\, \sqg
\eea

\hfill

where $d^d X\equiv \prod_{a=1}^{d} dX^{a}$. But we are in the
$X^{a}=\ve^a$ gauge, so both actions are off-shell the same :

\be
\label{d21}
S_{\phi}^{G}=\int_{\cal M}d^{D}X \sqp \, \pdt = \int_{\S}d^{d}\ve \sqg=
S_{p}^{G}
\ee

\hfill

This proof does not use any privileged choice of coordinates in the
target-space
but it uses a privileged one in the world-volume. However, 
any other
choice in the world-volume can be reached doing the target-space 
diffeomorphism $X^a \rightarrow X^{'a}(X^b)$ and $X^i \rightarrow X^i $
before solving the $\phi^i (X^{\m})=0$ equations. Then, the procedure
applies again to get (\ref{d21}). 


\section{The Electromagnetic Contribution}

In this section we will see how represent, in this dual
formulation, the electromagnetic contribution, i.e., the 
coupling with the d-form $\A$. 
In this case, the invariance under redefinitions of the fields,
diffeomorphismes in the target-space,
and the gauge invariance
$\A \rightarrow \A+ \d_{[\nu_{1}}\lambda_{\n_2 ,...,\n_{d}]}$,
provides the natural candidate for the action :

\be
\label{e1}
S^{A}_{\phi}={1\over d!}\int_{\cal M}d^D X \pdt \,\, 
\e^{\n_1 ,...,\n_{d}\m_1 ,...,\m_{\td}} \A \ppd
\ee

\hfill

Now, the d-form's contribution to the classical equations is

\be
\label{e6}
{\delta S_{\phi}^{A} \over \delta \phi^i}=
{1\over (d+1)!}F_{\m_{i} \n_{1},...,\n_{d}}
\e^{\n_{1},...,\n_{d}\m_{1},...,\m_{i},...,\m_{\td}}\ppdi
\ee

\hfill

To evaluate the implications of this contribution 
when restricted to the $\S$ hypersurface,i.e., the implications
for the embedding $X^{\m}(\ve^a)$,  we must take
account the closure relation derived from (\ref{d12})

\be
\label{e3}
\delta_{\m}^{\n}= P_{\mu}^{\n}+U^{\S\,\n}_{\m}=
h_{ij}\d_{\n}\phi^{i} \d^{\m}\phi^{j}+
\gamma^{ab}\d_{a}X^{\n}\d_{b}X^{\rho}G_{\rho\m}
\ee

\hfill

over the free indices in $\e^{\n_1 ,...,\n_{d}\m_1 ,...,\m,...,\m_{\td}}
F_{\m\n_{1},...,\n_{d}}$ and the kinematical relation
$J^2\,G=\gamma\,h$ on $\S$, with $J=det\{\d_{a}X^{\m}\, , \d^{\m}\phi^i\}$.
Finally we get :

\be
\label{eee}
({\delta S^{A}_{\phi}\over \delta \phi^i})_{\S}=
(-1)^{J}({G\,h\over\gamma})^{1\over 2}h_{ij}\d^{\m}\phi^j 
F_{\m\n_{1},...,\n_{d}}\pxd
\ee

\hfill

where $(-1)^{J}$ is the sign of the $J$ determinant.
Choosing the orientation of the $\d \phi^i $ vectors
in such a way that $(-1)^{J}=1$
and, then, 
joining equations (\ref{d13}), (\ref{d15}) and (\ref{eee})
we obtain one of the main results:

\be
\label{eee1}
({\delta S_{\phi}\over \delta \phi^i}\d_{\m}\phi^i)_{\S}
=({\delta (S_{\phi}^{G}+ S_{\phi}^{A})\over \delta \phi^i}
\d_{\m}\phi^i )_{\S}=
\ee

\be
\label{eeee11}
({G\,h\over\gamma})^{1\over 2}(- 
\sqrt{\gamma}P_{\n\m}^{\S}
\Delta X^{\m} + F_{\n\n_{1},...,\n_{d}}\pxd)=
\ee

\be
\label{eeeee111}
=-({G\,h\over\gamma})^{1\over 2}{\delta S_{p}\over
\delta X^{\n}}
\ee

\hfill

Therefore, the field equations for the scalar  fields, including the
electromagnetic
contribution, imply, when evaluated on the hypersurface where all the
fields vanish,
the p-brane equations for any embedding of that hypersurface. The converse
is also true. It is relevant to note that the classical scalar fields
couple
with the Hodge dual of the electromagnetic field strenght (\ref{e6}).
Then, the relation between both brane formulations is an 
electric-magnetic-duality.

\hfill

Following the same steps as in the preceding section, it can be shown 
that the electromagnetic contribution (\ref{e1}) reduces off-shell to
the p-brane electromagnetic action (\ref{p1bis}) : first, we solve 
the $\phi^i(X^\m)=0$ equations splitting the coordinates
in such a way that the solution is $X^{i}=X^{i}(X^{a})$.
Second, we choose the gauge $X^{a}=\ve^a$ in the world-volume
,third, we integrate in the transverse coordinates with the
net effect of giving a jacobian $\mid det({\d X^{i}\over
\d \phi^{j}}) \mid $ and the restriction $X^{i}=X^{i}(X^{a})$,
and  fourth, we use the closure relation (\ref{e3}) :

\bea
\label{e2}
&S_{\phi}^{A}=
{1\over d!}\int d^d X\,(\int \pdt dX^{i} 
\A \e^{\n_1 ,...,\n_{d}\m_1 ,...,\m_{\td}} 
\ppd)=\nonumber\\
&{1\over d!}\int d^d X 
\{\mid det({\d X^{i}\over
\d \phi^{j}})\mid 
\A \e^{\n_1 ,...,\n_{d}\m_1 ,...,\m_{\td}} 
\ppd\}\vert_{X^{i}=X^{i}(X^{a})}=\nonumber\\
& =\int d^d X \{\mid det({\d X^{i}\over
\d \phi^{j}})\mid {G\,J\over \gamma}  \A\}
\vert_{X^{i}=X^{i}(X^{a})}
\pxd
\eea

\hfill

and the kinematic identities, say, 
$\mid det({\d X^{i}\over \d \phi^{j}}) \mid {G\,J\over \gamma} =1$
\footnote{Choosing again the $(-1)^{J}=1$ orientation.},
providing the off-shell equality with the p-brane action :

\bea
\label{e5}
&S_{\phi}^{A}=\int_{\cal M}d^D X \pdt \,\, 
\e^{\n_1 ,...,\n_{d}\m_1 ,...,\m_{\td}} \A \ppd = \nonumber\\
&\int_{\S} d^d \ve  \A \pxd = S_{p}^{A}
\eea

\hfill

As a simple example of this formulation, we get the existence
of classical $(D-2)$-spherical-branes in D-dimensional 
euclidean flat space. Obviously,
this is possible thanks to the presence of a certain electromagnetic 
field :

\be
\label{e7}
\A= {1\over r \,(D-1) !}\e_{\n_{1},...,\n_{d} \m}X^{\m}
\ee

\hfill

with $X^{\m}$ cartesian coordinates and the radius 
$r\equiv (X^{\m} X^{\m})^{{1\over 2}}$.
This $(D-1)$-form allows the existence of the spherical solution 
$\phi = \phi (X^{\m}X ^{\m})$ (here $\td =1$). The electromagnetic
contribution compensates the non null one coming from the
coupling with the flat metric :

\be
\label{e8}
{\delta \sqrt{G\,h} \over \delta \phi} =
{(D-1)\over r}
\ee

\hfill

As we can see, (\ref{e8}) is independent of the functional form
$\phi (r^2)$. Finally, $\phi (X^{\m}X ^{\m})=0$ describes the
$(D-1)$-sphere $r^2=r_{0}^2$, now becoming a classic $(D-2)$-brane
in flat space.
\section{Redefinition invariance and branes}
The purpose of this section is to exhibit how the brane emerges
in a natural way from the field theory, just due to the invariance under 
the redefinition of the fields (\ref{d2}). In this context, the 
reparametrization invariance in the world-volume, exists as a result 
of the general covariance under target-space diffeomorphisms.

\hfill

Let us start with the most general action for the set of  $\td$ scalar
fields
$\phi^i(X)$, with invariance under field redefinitions
$\phi^i(X)\rightarrow
\phi^{'i}(\phi^{j}(X))$ and being a candidate to describe an 
elementary extended object $\phi^i(X)=0$ :

\be
\label{gr1}
S_{\phi}=\int_{M}d^D X \pdt \,{\cal L}_{\phi}
\ee

\hfill

The restriction over $\phi^i = 0$ reduces the $\phi$-dependence in 
the lagrangian density to be ${\cal L}_{\phi}={\cal L}(\d_{\m}\phi^i)$. 
Moreover, the invariance of the action under field redefinitions 
forces the transformation 

\be
\label{gr2}
{\cal L}_{\phi '}=det({\d \phi^{'}\over \d \phi}){\cal L}_{\phi}
\ee

\hfill 

because of the delta functions transformation (\ref{d19}). 
As was proved  in \cite{fairlie}, is just this covariance which guarantees
the 
covariance of the field equations derived from that lagrangian. 

\hfill

Let us now assume that we solve the algebraic equations $\phi^i(X)=0$
,defining the hypersurface $\S$, in arbitrary coordinates through the
split $\{X^i , X^a\}$ with $a=\td+1$ to $D$, in the way $X^i=X^i(X^a)$.
Choosing the gauge $X^a=\ve^a$ in the world-volume, we have seen 
that (\ref{d10}) explicitly gives 
$(\d_{a}\phi^i)_{\S}=-\d_{a}X^{j}(\d_{j}\phi^i)_{\S}$.
Using that, we can rewrite 
$(\d_{\m}\phi^i)_{\S}=M_{j}^{i}\d_{\m}\bar{\phi}^i$,
where the barred field configuration is $\bar{\phi}^i=X^i-X^i(X^a)$
\footnote{It must be stressed that, in general, this barred configuration
cannot
reached by field redefinitions, although it seems to be the simplest one
describing the extended object. If we try to introduce the barred 
configuration into the field equations in, for example, (\ref{d8}),
we see that the equations are inconsistent, except in the $X^i=X^i(X^a)$
hypersurface in which case they become the p-brane equations in the 
$X^a=\ve^a$ gauge. As we will see in this  section, this is no a
coincidence,
but a consecuence of the invariance under field redefinitions.}
and the matrix $M^i_j={\d \phi^i \over \d X^{j}}$. The point is that 
(\ref{gr2}) implies ${\cal L}_{\phi}(M^{i}_{j}\d_{\m}\phi^j)=
detM{\cal L}_{\phi}(\d_{\m}\phi^i)$, for arbitrary matrix $M_i^j$,so we
have :

\be
\label{gr3}
S_{\phi}=\int_{M}d^D X \pdt \,( det {\d \phi^i \over \d X^j})
{\cal L}_{\phi}(-\d_{a}X^i , \delta_j^i)
\ee

\hfill

but $ \mid det {\d \phi^k \over \d
X^j}\mid \pdt=\prod_{i=1}^{\td}\delta(X^i-X^i(X^a))$
\footnote{We have seen in the last section that there is an 
orientation which must be choosen. This possibility is ruled out
when the lagrangian transforms as 
${\cal L}_{\phi '}=\mid det({\d \phi^{'}\over \d \phi}) \mid 
{\cal L}_{\phi}$, as in  (\ref{d16}).}
, and we can integrate in the $X^i$ transverse coordinates, letting 
an action in which the explicit dependence on the scalar fields 
has totally disappeared in favor of the embedding's $X^{\m}(\ve^{a})$
dependence, in the gauge $X^a=\ve^a$. Therefore, the starting action
$S_{\phi}$ can be written, in a general way, only in terms of the 
embedding of the hypersurface 

\be
\label{gr4}
S_{\phi}=\int_{\S}d^d X {\cal L}_{\phi}(-\d_{a}X^i , \delta_j^i)\equiv
\int_{\S}d^d \ve {\cal L}_{X} = S_{X}
\ee

\hfill

or, in other words, what we have done is to obtain the world-volume 
gauge fixed brane lagrangian ${\cal L}_{X}$ from the target-space one 
${\cal L}_{\phi}$, in the way :

\bea
\label{gr5}
&{\cal L}_{X}(X^{a}=\ve^{a}, \d_{a}X^i )= 
{\cal L}_{\phi}(\phi^i = \bar{\phi}^i )=\nonumber\\
&={\cal L}_{\phi}(\d_{a}\phi^i=-\d_{a}X^i , \d_{j}\phi^i =\delta^i_j )
\eea

\hfill

and that works only with the help of the invariance under redefinitions, it
doesn't matter the particular details of the $\phi$-dependence. Those
details 
are, of course, relevant if we are interested in knowing what brane theory
we are dealing with, as it is the case in this paper. Moreover, the
invariance
of the starting action under target-space diffeomorphisms ensures that 
${\cal L}_{X}$ is the gauged-fixed action of a truly target-space and
world-volume invariant one.

\hfill

More important to us is the converse problem, i.e., the derivation of the
target-space field theory once the brane one is already given. From the
above considerations, it is easy to realize that the invariance under field
redefinitions solves again the problem. In our split $\{X^i , X^a \}$, 
we can write :

\bea
\label{gr6}
&{\cal L}_{\phi}(\d_{\m}\phi^i) = 
{\cal L}_{\phi}(\d_{a}\phi^i , \d_{j}\phi^i )=
\nonumber\\
& ( det {\d \phi^i \over \d X^j}){\cal L}_{\phi}
({\d X^i \over \d \phi^j}{\d \phi^j \over \d X^a}, \delta_{j}^{i})
\eea

\hfill

But using (\ref{gr5}) we get the final result :

\be
\label{gr7}
{\cal L}_{\phi}(\d_{\m}\phi^i)= ( det {\d \phi^i \over \d X^j})
{\cal L}_{X}(\ve^a = X^a , \d_{a}X^{i}=
-{\d X^i \over \d \phi^j}{\d \phi^j \over \d X^a})
\ee

\hfill

Of course, the split  of coordinates $\{X^a , X^i \}$ breaks 
the manifest covariance under target-space diffeomorphismes, but
if the original brane lagrangian is properly built, this breaking
is only apparent. Let's do a very simple example ; the coupling
between the 0-brane and the electromagnetic field is given by the
term ${\cal L}_{X}^{A}=A_{\m}(X(\tau))\dot{X}^{\m}= 
A_{0}(\tau, X^k (\tau) ) + A_{i}(\tau, X^j (\tau) ) \dot{X}^i $,
which is already written in the gauge $X^0 = \tau$ and with
$\dot{X}^{\m}={d X^{\m}\over d\tau}$. Following the 
prescription (\ref{gr7}) the corresponding term in the target-space
field theory will be :

\be
\label{gr8}
{\cal L}^{A}_{\phi}(\d_{\m}\phi^i)= ( det {\d \phi^i \over \d X^j})(
A_{0}(X^0, X^k ) - A_{i}(X^0, X^j)
{\d X^i \over \d \phi^j}{\d \phi^j \over \d X^0})
\ee

\hfill

which, in fact, has the target-space manifest covariant form :

\be
\label{gr9}
{\cal L}^{A}_{\phi}(\d_{\m}\phi^i)=A_{\m}(X)\epsilon^{\m,
\n_{1}...\n_{D-1}}
\d_{\n_{1}}\phi^1 \times ... \times \d_{\n_{D-1}}\phi^{D-1}
\ee

\hfill

Of course, this example agrees with (\ref{e1}) in the case $d=1$. 
Even more, the formula (\ref{gr7}) will be used in the next section 
to get the D-brane dual target-space lagrangian. 

\hfill

So far, we have elucidated the relation between the lagrangians
for both formulations of the same extended object. To end this section,
we give the explicit formula relating both dynamics, and again, 
obtained with the extensive help of the invariance under field 
redefinitions, i.e., via (\ref{gr2}) and its implications.

Given the lagrangian for our field theory ${\cal L}_{\phi}$ 
satisfying the covariance transformation (\ref{gr2})
under field redefinitions, and given the gauge fixed brane  
lagrangian ${\cal L}_{X}$ in terms of it via (\ref{gr5}), 
it can be shown that the field 
equations for both theories are related by the expression:

\be
\label{gr10} 
({\delta S_{\phi} \over \delta \phi^i})_{\S}=
- (det({\d \phi \over \d X}){\d X^j \over \d \phi^i})_{\S}
{\delta S_{X}\over \delta X^j }
\ee

\hfill

If the target-space action is properly built
(invariance under space-time diffeomorphisms and 
of course, field redefinitions), the brane theory
built from it through (\ref{gr5})
will have the correspondent invariance
under world-volume reparametrizations. In that
case, the ${\delta S_{X}
\over \delta X^i }=0$ equations imply the remaining 
${\delta S_{X}\over \delta X^a }=0$ ones,
due to the identity
 ${\delta S_{X} \over \delta X^{\m}}{\d X^{\m}\over \d \ve^a}=0$,
forced by the reparametrization invariance.
This identity allow us to write
the on-shell equivalence relation (\ref{gr10})
in the final full covariant form:

\be
\label{gr11}
({1\over {\cal L}_{\phi}}
{\delta S_{\phi} \over \delta \phi^i}
\d_{\m}\phi^i )_{\S}=
-{1\over {\cal L}_{X}}{\delta S_{X}\over
\delta X^{\m}}
\ee

\hfill

This identity was explicitily obtained in the
the last sections (\ref{eee1})
for the p-branes. This was made without
fixing any gauge in the world-volume,
but paying attention to the particular form of the 
actions we worked with.
Now the formula (\ref{gr11}) was derived 
just with the help of the invariance under field 
redefinitions, no matter the specific for of the action 
is. From the point of view of this general approach,
the only work we should do in trying to get the 
dual version of a brane theory, is to apply the formula
(\ref{gr7}) and then (\ref{gr11}) satisfies identically,
providing the classical equivalence.

\section{The D-brane's dual model}
After the detailed calculations for the p-branes 
have been reproduced in the previous sections, 
and the general features of the world-volume/target-space
dualization procedure have been elucidated in the last section,
a few remarks concern for the Dirichlet-branes. 
The formula (\ref{gr7}) gives the target-space action 
in terms of the DBI one, which results 
to be the (\ref{d18}) action with the replacement
$G_{\m\n}\rightarrow (G+B)_{\m\n}\equiv Q_{\m\n}$.
The dual version of the electromagnetic contribution is the same 
as in the fifth section. The only difference, irrelevant to this
dualization
procedure, is that the d-form is given by a certain combination of 
the R-R forms and the NS-NS two-form. 

\hfill

To organize our results in such a way that the covariance 
under field redefinitions is manifest, we follow the same 
line as in the p-brane case.  The main difference is the technical 
complications just due to the NS-NS two form $B_{\m\n}$ in the DBI-action 
and in the (\ref{d18})-type action. Now, the natural first-derivative
target-space scalar (\ref{d3}) has an antisymmetric part, which is
encoded in :

\be
\label{dd3}
h^{ij}_{\pm}(X) = Q(X)^{\m\n}_{\pm}\ppi \ppj
\ee 

\hfill

where we define $Q_{\m\n}^{\pm}\equiv (G\pm B)_{\m\n}$,
and 
$Q^{\m\alpha}_{\pm}Q_{\alpha\n}^{\pm}=\delta_{\n}^{\m}$.
(\ref{dd3}) imply the existence of two new privileged conections
under field redefinitions 
\footnote{$h_{ik}^{\pm}h^{kj}_{\pm}=\delta_{i}^{j}$ and
the target-space covariant derivative $\D_{\m}$ continues being 
constructed with the Levi-Civita connection}:

\be
\label{dd5} 
(\O_{\m}^{\pm})_{i}^{j} \equiv \O_{\m\,\,i}^{\pm\,j} =
\D_{\m}\d_{\n}\phi^{j} Q_{\mp}^{\n\rho}\d_{\rho}\phi^{k} h_{ki}^{\mp}
\ee

\hfill

with the corresponding covariant derivations :

\bea
\label{dd62}
&V^{'i}=  {\d \phi^{' i}\over \d \phi^{k}} V^{k} \rightarrow
D_{\m}^{\pm}V^{i}\equiv \d_{\m}V^{i}-\O_{\m\,j}^{\pm\,i}V^{j}\rightarrow
D_{\m}^{\pm '}V^{'i}= {\d \phi^{' i}\over \d \phi^{k}}
D_{\m}^{\pm}V^{k}\nonumber\\
&V^{'}_{i}=  {\d \phi^{ j}\over \d \phi^{'i}} V_{j}
\rightarrow D_{\m}^{\pm}V_{i}\equiv \d_{\m}V_{i}+ \O_{\m\,i}^{\pm\,j}V_{j}
\rightarrow D_{\m}^{\pm '}V_{i}^{'}= {\d \phi^{ j}\over \d \phi^{'i}}
D_{\m}^{\pm}V_{j}
\eea

\hfill

But this plurality of connections prevents the existence of
a privileged dynamic. Therefore, the straight way is 
to use the formula (\ref{gr7}) giving the lagrangian for the 
target-space scalar fields in terms of the D-brane one contained
in the DBI-action :

\be
\label{ppp1}
S_{DBI}^{Q}= \int_{\S} d^d \ve  \sqrt{\g^{+}}
\ee

\hfill

where $\g_{ab}^{\pm}=Q_{\m\n}^{\pm}(X)\d_{a}X^{\m}\d_{b}X^{\n}$
is the induced tensor and $\gamma^{+}$ its determinant.
The result is the generalization of the action (\ref{d18}),

\be
\label{dd18}
S_{\phi}^{Q}=\int_{\cal M}d^{D}X \,\sqqp \,\pdt 
\ee

\hfill

where $Q^{+}\equiv det\,Q_{\m\n}^{+}=det\,Q_{\m\n}$ and 
$h^{+}\equiv det \,h^{ij}_{+}$.
The field equations are :

\bea
\label{dd35}
&{\delta S_{\phi}^{Q}\over \delta \phi^i}\d_{\m}\phi^{i}=
\sqqp \{\d_{\m}\phi^{i}{1\over2}(h^{+}+h^{-})_{ij}Q_{+}^{\alpha\beta}
D_{\alpha}^{+}\d_{\beta}\phi^j-\nonumber\\
&-{1\over 4}(U^{\rho}_{\,\,\m}-U^{\,\,\rho}_{\m}) 
(\O_{\rho\,k}^{+ \,k}-\O_{\rho\,k}^{- \,k})+ \nonumber\\
&+U^{\lambda\rho}(-H_{\rho\lambda\m}+
{1\over2}U_{\m}^{\,\,\sigma}(\D_{\rho}B_{\sigma\lambda}-
{1\over2}\D_{\sigma}B_{\rho\lambda})+\nonumber\\
&{1\over2}U_{\,\,\m}^{\sigma}(\D_{\lambda}B_{\rho\sigma}-
{1\over2}\D_{\sigma}B_{\rho\lambda}))\}=0
\eea

\hfill

\footnote{$H_{\m\n\rho}\equiv {1\over2}(
\d_{\m}B_{\n\rho}+\d_{\n}B_{\rho\m}+\d_{\rho}B_{\m\n}
)$.}
; they are written manifestly invariant under field redefinitions
with the help of the connections (\ref{dd5}) and 
the two different projectors $U_{\m}^{\,\,\n}$ and $ U^{\n}_{\,\,\m}$,
both coming from 
$U_{\m\n}\equiv Q^{+}_{\m\n}-h_{ij}^{+}\d_{\m}\phi^i \d_{\n}\phi^j$,
in the way $U_{\m}^{\,\,\n} =U_{\m\rho}Q^{\rho\n}_{+}$ and
$U^{\n}_{\,\,\m}= Q^{\n\rho}_{+}U_{\rho\m}$. 
The variational derivative (\ref{dd35}) is ready for a direct comparison
with the equations for the D-brane's embedding coming from the 
DBI-action \footnote{See Appendix}. The main result is again the 
identiy :

\be
\label{eee11}
({1\over\sqrt{Q^{+} h^{+}}}{\delta S_{\phi}^{Q}\over \delta 
\phi^i}\d_{\m}\phi^{i})_{\S} = -{1\over\sqrt{\gamma^{+}}} 
{\delta S_{DBI}^{Q}\over
\delta X^{\m}}
\ee

\hfill

on $\S$ providing the on-shell equivalence of both formulations.
Finally, the construction via the formula (\ref{gr7}) trivially implies 
the off-shell identity between the DBI-action and the (\ref{dd18}) one. 

\be
\label{dd11}
S_{\phi}^{Q}=S_{DBI}^{Q}
\ee
 
\hfill

Again, because of the invariance under global translations of the 
field equations (\ref{dd35}), every solution $\phi^i(X)$
splits the whole space-time into a continuous of D-branes, a  fluid 
of  classic D-branes. 
\section{Conclusions}
In this work, the equations for a set of target-space
scalar fields $\phi^i (X^\m)$ for $i=1\,$to$\,
\td$, are built in such a way that they have a direct
interpretation as representing a continuous of d-dimensional
($d+\td=D$) extended objects 
$\S_{\phi_ 0}$,everyone of them implicitly defined as the
collection of the target-space points where the scalar
fields adquire a given constant value, i.e.,
through the condition $\phi^i (X^\m)=\phi^{i}_{0}$.
The fundamental criteria for getting the dynamic and
relating it with branes, is the invariance under field
redefinitions of the scalar fields.   
Even more, it is shown in detail how these extended objects are 
classical p-branes or D-branes, depending on what the field 
theory it is, then, 
fully supporting the construction.
After the proof of the on-shell equivalence with p(or D)-branes 
have been made,
the right actions providing the field equations and transforming
nicely under the field redefinitions are found. 
These actions
correspond to an elementary extended object instead of
a continuous of them, and it is shown how it reduces off-shell
(kinematically) to the p-brane and D-brane 
actions respectively. Therefore 
they are dual formulations of  p(or D)-branes.
The classical equivalence is of the electric-magnetic duality-type
, because 
the dual fields couple classically to the dual (in the Hodge
sense) field strength. The elementary feauture of the actions
makes harmless the non-uniqueness of the lagrangian density,
then giving a sense for this ambiguity, at the same time that
it is the responsible for the invariance of the action. 

\hfill

Moreover, it is shown how the elementarity and the invariance
under field redefinitions are the basic properties allowing
the generic exchange between the target-space $\phi$-degree of freedom
and the embedding ones in the world-volume.
The explicit expressions relating the lagrangians (and the dynamics)
of both, the target-space and the world volume theories are worked 
out just with the help of the invariance under field redefinitions, which
results to be the basic ingredient of this dualization procedure. 

All this questions are solved in arbitrary space-time
geometry, dimension, electromagnetic d-form field and dimensionallity
of the brane.

{\bf Acknowledgements.}
I am grateful to Enrique Alvarez,  C\' esar G\'omez and Pedro Silva
by their comments. This work was supported by a
Comunidad Aut\'onoma de Madrid grant.
\appendix
\section{Appendix : The D-brane Equations}

In this appendix we are interested on getting the equations
for the D-brane's embedding. The main difference with the
p-brane case lies in the presence of the NS-NS two-form.Then,
we will only evaluate the kinetic part coming from the DBI-action 
defined in the last section (\ref{ppp1}). The task is to put in order
the equations following two criteria : the explicit world-volume and 
target-space covariance, and our intention to compare with the
field equations (\ref{dd35}). In order to understand the covariance 
structure of the equations let us start  looking over the p-brane ones.
Given the embedding $X^{\m}(\ve^a)$ of a hypersurface $\S$
defined in a Riemannian manifold $M$, we can always define
the pullback of the (compatible with the metric) connection\footnote{The
Levi-Civita one through this paper.} as 
$\Gamma_{a\m}^{\n}\equiv \d_{a}X^{\rho}\Gamma_{\rho\m}^{\n}$,
and the world-volume connection as $\gamma_{ab}^{c}\equiv
\gamma^{cd}\d_{d}X^{\n}G_{\n\m}\D_{a}\d_{b}X^{\m}$, where
$\gamma_{ab}$ is the induced metric and $\D_{a}\d_{b}X^{\m}=
\d_{a}\d_{b}X^{\m}-\Gamma_{\alpha\beta}^{\m}\d_{a}X^{\alpha}
\d_{b}X^{\beta}$ is the target-space (although no world-volume)
embedding's covariant derivative\footnote{It can be checked that the
world-volume connection coincides with the Levi-Civita one of 
the induced metric $\gamma$ always than the target-space one be
the Levi-Civita connection.}. Then, a simple way to prescribe a
target-space
and world-volume covariant dynamic for the embedding is through :

\be
\label{ap1}
\gamma^{ab}D_{a}\d_{b}X^{\m}\equiv
\gamma^{ab}(\d_{a}\d_{b}X^{\m}+ \Gamma_{a\n}^{\m}\d_{b}X^{\n}
-\gamma_{ab}^{c}\d_{c}X^{\m})=0
\ee

\hfill

which, after a simple calculation results to be the p-brane equation
without
electromagnetic field (\ref{p2}),i.e.,  
$\gamma^{ab}D_{a}\d_{b}X^{\m}=\gamma^{ab}
P^{\m}_{\,\,\n}\D_{a}\d_{b}X^{\n}$.

\hfill

Now, the NS-NS two-form allows two new different possibilities
for a world-volume connection \footnote{As it happended in the 
target-space field theoretic dual version, (\ref{dd5}).} :

\be
\label{ap2}
\gamma_{ab}^{\pm c}\equiv
\gamma^{cd}_{\pm}\d_{d}X^{\n}Q_{\n\m}^{\pm}\D_{a}\d_{b}X^{\m} 
\ee

\hfill

and the corresponding target-space/world-volume covariant derivations :

\be
\label{ap3}
D_{a}^{\pm}\d_{b}X^{\m}\equiv
\d_{a}\d_{b}X^{\m}+ \Gamma_{a\n}^{\m}\d_{b}X^{\n}
-\gamma_{ab}^{\pm c}\d_{c}X^{\m}
\ee

\hfill

Now we are ready to structurate 
the D-brane equations, which, moreover will be specially
adapted to compare with the field equations (\ref{dd35}) when
evaluated on $\S$ :

\bea
\label{ap4}
&{\delta S_{DBI}^{Q}\over \delta X^{\m}}=
\sqrt{\gamma^+} \{{1\over2}(\gamma^{ab}_{+}
Q_{\m\alpha}^{+}D_{a}^{+}\d_{b}
X^{\alpha}+\gamma^{ab}_{-}Q_{\m\alpha}^{-}D_{a}^{-}\d_{b}
X^{\alpha})-\nonumber\\
&-{1\over 4}(\gamma^{ab}_{-}\d_{b}X^{\lambda}Q_{\lambda\m}^{-}-
\gamma^{ab}_{+}\d_{b}X^{\lambda}Q_{\lambda\m}^{+}) 
(\gamma_{ac}^{-c}-\gamma_{ac}^{+c})+ \nonumber\\
&+U^{\lambda\rho}_{\S}(H_{\rho\lambda\m}-
{1\over2}U_{\m}^{\S\,\sigma}(\D_{\rho}B_{\sigma\lambda}-
{1\over2}\D_{\sigma}B_{\rho\lambda})-\nonumber\\
&-{1\over2}U_{\S\,\m}^{\sigma}(\D_{\lambda}B_{\rho\sigma}-
{1\over2}\D_{\sigma}B_{\rho\lambda}))\}=0
\eea
 
\hfill

where now we have two projectors built from
 $U^{\m\n}_{\S}\equiv  \g^{ab}_{+}
\d_{a}X^{\m}\d_{b}X^{\n}$ in the way
$U^{\n}_{\S\,\m}=U^{\n\rho}_{\S} Q_{\rho\m}^{+}$ and
$U^{\S\,\n}_{\m}= Q_{\m\rho}^{+}U^{\rho\n}_{\S}$.
They act as the identity over the brane directions, but they define 
two different "ortogonal"
directions, i.e., $ \d_{\beta}\phi^{i} Q^{\beta\n}$ and $ Q^{\n\beta}
\d_{\beta}\phi^{i}$. Of course, the relation between the D-brane
equations (\ref{ap4}) and the field equations (\ref{dd35}), 
uses again the identity on $\S$, $ U^{\m\n}_{\S}=(U^{\m\n})\vert_{\S}$
where $U^{\m\n}=Q^{\m\alpha}_{+}Q^{\n\beta}_{-}U_{\alpha\beta}$
getting the result (\ref{eee11}).


\end{document}